\documentclass[11pt,fleqn]{article}
\usepackage{amsfonts, epsfig, amssymb, amsmath}
\usepackage[normalem]{ulem}
\usepackage{color}
\newcommand{\ol}{\setlength{\itemsep}{0pt.}\begin{enumerate}}
\newcommand{\eol}{\end{enumerate}\setlength{\itemsep}{-\parsep}}
\newcommand{\ignore}[1]{}
\title{On some properties of random and pseudorandom codes}
\author{Alex Samorodnitsky}

\begin{document}
\date{}
\maketitle


\newtheorem{THEOREM}{Theorem}[section]
\newenvironment{theorem}{\begin{THEOREM} \hspace{-.85em} {\bf :}
}%
                        {\end{THEOREM}}
\newtheorem{LEMMA}[THEOREM]{Lemma}
\newenvironment{lemma}{\begin{LEMMA} \hspace{-.85em} {\bf :} }%
                      {\end{LEMMA}}
\newtheorem{COROLLARY}[THEOREM]{Corollary}
\newenvironment{corollary}{\begin{COROLLARY} \hspace{-.85em} {\bf
:} }%
                          {\end{COROLLARY}}
\newtheorem{PROPOSITION}[THEOREM]{Proposition}
\newenvironment{proposition}{\begin{PROPOSITION} \hspace{-.85em}
{\bf :} }%
                            {\end{PROPOSITION}}
\newtheorem{DEFINITION}[THEOREM]{Definition}
\newenvironment{definition}{\begin{DEFINITION} \hspace{-.85em} {\bf
:} \rm}%
                            {\end{DEFINITION}}
\newtheorem{EXAMPLE}[THEOREM]{Example}
\newenvironment{example}{\begin{EXAMPLE} \hspace{-.85em} {\bf :}
\rm}%
                            {\end{EXAMPLE}}
\newtheorem{CONJECTURE}[THEOREM]{Conjecture}
\newenvironment{conjecture}{\begin{CONJECTURE} \hspace{-.85em}
{\bf :} \rm}%
                            {\end{CONJECTURE}}
\newtheorem{MAINCONJECTURE}[THEOREM]{Main Conjecture}
\newenvironment{mainconjecture}{\begin{MAINCONJECTURE} \hspace{-.85em}
{\bf :} \rm}%
                            {\end{MAINCONJECTURE}}
\newtheorem{PROBLEM}[THEOREM]{Problem}
\newenvironment{problem}{\begin{PROBLEM} \hspace{-.85em} {\bf :}
\rm}%
                            {\end{PROBLEM}}
\newtheorem{QUESTION}[THEOREM]{Question}
\newenvironment{question}{\begin{QUESTION} \hspace{-.85em} {\bf :}
\rm}%
                            {\end{QUESTION}}
\newtheorem{REMARK}[THEOREM]{Remark}
\newenvironment{remark}{\begin{REMARK} \hspace{-.85em} {\bf :}
\rm}%
                            {\end{REMARK}}

\newcommand{\thm}{\begin{theorem}}
\newcommand{\lem}{\begin{lemma}}
\newcommand{\pro}{\begin{proposition}}
\newcommand{\dfn}{\begin{definition}}
\newcommand{\rem}{\begin{remark}}
\newcommand{\xam}{\begin{example}}
\newcommand{\cnj}{\begin{conjecture}}
\newcommand{\mcnj}{\begin{mainconjecture}}
\newcommand{\prb}{\begin{problem}}
\newcommand{\que}{\begin{question}}
\newcommand{\cor}{\begin{corollary}}
\newcommand{\prf}{\noindent{\bf Proof:} }
\newcommand{\ethm}{\end{theorem}}
\newcommand{\elem}{\end{lemma}}
\newcommand{\epro}{\end{proposition}}
\newcommand{\edfn}{\bbox\end{definition}}
\newcommand{\erem}{\bbox\end{remark}}
\newcommand{\exam}{\bbox\end{example}}
\newcommand{\ecnj}{\bbox\end{conjecture}}
\newcommand{\emcnj}{\bbox\end{mainconjecture}}
\newcommand{\eprb}{\bbox\end{problem}}
\newcommand{\eque}{\bbox\end{question}}
\newcommand{\ecor}{\end{corollary}}
\newcommand{\eprf}{\bbox}
\newcommand{\beqn}{\begin{equation}}
\newcommand{\eeqn}{\end{equation}}
\newcommand{\wbox}{\mbox{$\sqcap$\llap{$\sqcup$}}}
\newcommand{\bbox}{\vrule height7pt width4pt depth1pt}
\newcommand{\qed}{\bbox}
\def\sup{^}

\def\H{\{0,1\}^n}

\def\S{S(n,w)}

\def\g{g_{\ast}}
\def\xop{x_{\ast}}
\def\y{y_{\ast}}
\def\z{z_{\ast}}

\def\f{\tilde f}

\def\n{\lfloor \frac n2 \rfloor}

\def \E{\mathop{{}\mathbb E}}
\def \R{\mathbb R}
\def \Z{\mathbb Z}
\def \F{\mathbb F}
\def \T{\mathbb T}

\def \x{\textcolor{red}{x}}
\def \r{\textcolor{red}{r}}
\def \Rc{\textcolor{red}{R}}

\def \noi{{\noindent}}

\def \iff{~~~~\Longleftrightarrow~~~~}

\def\myblt{\noi --\, }

\def \queq {\quad = \quad}

\def\<{\left<}
\def\>{\right>}
\def \({\left(}
\def \){\right)}

\def \e{\epsilon}
\def \l{\lambda}

\def\Tp{Tchebyshef polynomial}
\def\Tps{TchebysDeto be the maximafine $A(n,d)$ l size of a code with distance $d$hef polynomials}
\newcommand{\rarrow}{\rightarrow}

\newcommand{\larrow}{\leftarrow}

\overfullrule=0pt
\def\setof#1{\lbrace #1 \rbrace}

\begin{abstract}
We describe some pseudorandom properties of binary linear codes achieving capacity on the binary erasure channel under bit-MAP decoding (as shown in \cite{KKMPSU} this includes doubly transitive codes and, in particular, Reed-Muller codes). We show that for all integer $q \ge 2$ the $\ell_q$ norm of the characteristic function of such 'pseudorandom' code decreases as fast as that of any code of the same rate (and equally fast as that of a random code) under the action of the noise operator. In information-theoretic terms this means that the $q^{th}$ R\'enyi entropy of this code increases as fast as possible over the binary symmetric channel. In particular (taking $q = \infty$) this shows that such codes have the smallest asymptotic undetected error probability (equal to that of a random code) over the BSC, for a certain range of parameters.

We also study the number of times a certain local pattern, a 'rhombic' $4$-tuple of codewords, appears in a linear code, and show that for a certain range of parameters this number for pseudorandom codes is similar to that for a random code.

\end{abstract}

\section{Introduction}

\noi We study properties of binary linear codes achieving capacity on the binary erasure channel (BEC) under bit-MAP decoding.

\noi To define such codes, we recall some notions and introduce notation. Recall that the rate of a binary linear code is defined as $R(C) = \frac{1}{n} \log_2(|C|) = \frac{\textup{dim}(C)}{n}$. Given a linear code $C$, let $r_C(\cdot)$ be the rank function of the binary matroid on $\{1,...,n\}$ defined by a generating matrix of $C$. (That is, $r_C(T)$ is the rank of the column submatrix of a generating matrix of $C$ which contains columns indexed by $T$.) Let $T \sim \l$ denote a random subset $T$ of $[n]$ in which each element is chosen independently with probability $0 \le \l \le 1$.

\dfn
\label{dfn:Becca}
A code $C \subseteq \H$ of rate $R$ (more precisely a family of codes indexed by $n$) achieves capacity under bit-MAP decoding for the BEC if
\[
\E_{T \sim R} \big(|T| - r_C(T)\big) ~\le~ o(n).
\]
In other words, let $v_1,...,v_n \subseteq \{0,1\}^{Rn}$ be the columns of a generating matrix of $C$. Then almost all subsets of $\{v_1,...,v_n\}$ of cardinality $Rn$ have almost full rank.
\edfn

\noi This definition coincides with information-theoretic definitions of this property (as stated, somewhat implicitly, e.g., in \cite{KKMPSU}). In the context of this paper we will often refer to such codes as 'pseudorandom'. This can be justified by the well-known (and easy to see) fact that random linear codes satisfy Definition~\ref{dfn:Becca} with high probability. It should be noted that, as shown in \cite{KKMPSU}, doubly transitive codes and, in particular, Reed-Muller codes are pseudorandom in the sense of Definition~\ref{dfn:Becca}. For future use we note that $C$ is pseudorandom iff its dual $C^{\perp}$ is pseudorandom (this is an easy consequence of the following fact: for any $S \subseteq [n]$ holds $r_{C^{\perp}}(S) = |S| + r_C\(S^c\) - Rn$).

\noi We study pseudorandom codes with the assistance of an inequality \cite{S-old} which quantifies the decrease in the $\ell_q$ norm of a function on the boolean cube under noise. Before stating the claim of this inequality, we introduce the appropriate notion of a noise operator. Given a noise parameter $0 \le \e \le 1/2$, the noise operator $T_{\e}$ acts on functions on the boolean cube as follows: for $f:~\H \rarrow \R$, $T_{\e} f$ at a point $x$ is the expected value of $f$ at $y$, where $y$ is a random binary vector whose $i^{\small{th}}$ coordinate is $x_i$ with probability $1-\e$ and $1 - x_i$ with probability $\e$, independently for different coordinates. Namely, $\(T_{\e} f\)(x) =  \sum_{y \in \H} \e^{|y - x|}  (1-\e)^{n - |y-x|}  f(y)$, where $|\cdot|$ denotes the Hamming distance. We will write $f_{\e}$ for $T_{\e} f$, for brevity. We need one more item of notation. Given a subset $T \subseteq [n]$, let $\E(f|T)$ be the conditional expectation of $f$ with respect to $T$, that is a function on $\{0,1\}^n$ defined by $\E(f|T)(x) = \E_{y: y_{|T} = x_{|T}} f(y)$.

\thm
\label{thm:better-RE}
For any integer $q \ge 2$, and for any nonnegative function $f$ on $\H$ holds
\[
\log \|f_{\e}\|_q ~\le~ \E_{T \sim \l} \log \| \E(f|T)\|_q,
\]
with $\l = \l(q,\e) = 1 + \frac{1}{q-1} \cdot \log_2\(\e^q + (1-\e)^q\)$.

\noi We also have
\[
\log \|f_{\e}\|_{\infty} ~\le~ \E_{T \sim \l} \log \| \E(f|T)\|_{\infty},
\]
with $\l = \l(\infty,\e) = 1 +  \log_2\(1-\e\)$.
\ethm

\noi \cite{S, S-old, HSS} studied pseudorandom codes using the case $q=2$ of this inequality. In this paper we describe some simple consequences of this inequality for larger values of $q$, focusing in particular on $q=4$ and $q = \infty$.

\noi {\it Fast decrease of $\ell_q$ norms under noise}

\noi We start with the observation that the $\ell_q$ norms of the characteristic function of a pseudorandom code decrease as fast as possible under noise. Here and below we use $\approx$, $\gtrsim$, and $\lesssim$ notation to describe equalities and inequalities which hold up to lower order terms.

\pro
\label{pro:renyi entropy decrease}

\begin{itemize}

\item Let $C$ be a pseudorandom code of rate $0 < R < 1$, and let $f = \frac{2^n}{C} \cdot 1_C$. Then for any integer $q \ge 2$ and for any $0 \le \e \le \frac12$ such that $\l = \l(q,\e) = 1 + \frac{1}{q-1} \cdot \log_2\(\e^q + (1-\e)^q\) \ge R$ holds
\[
\frac 1n \log_2 \|f_{\e}\|_q ~\lesssim~ \frac{q-1}{q} \cdot (\l - R).
\]

\item Let $f$ be a nonnegative function on $\H$ with $\frac 1n \log_2 \|f\|_q = \frac{q-1}{q} \cdot (1-R)$. Then for any $q \ge 1$ and for any $0 \le \e \le \frac12$ holds
\[
\frac 1n \log_2 \|f_{\e}\|_q ~\ge~ \frac{q-1}{q} \cdot (\l - R).
\]

\end{itemize}

\noi Both claims also hold for $q = \infty$ (replacing $\frac{q-1}{q}$ by $1$) with $\l = \l(\infty,\e) = 1 +  \log_2\(1-\e\)$.

\epro

\noi Let us make two comments about this result.

\myblt Recall the notion of R\'enyi entropy from information theory: For a probability distribution $P$ on $\Omega$, the $q^{\small{th}}$ Renyi entropy of $P$ is given by $H_q(P) = -\frac{1}{q-1} \log_2\(\sum_{\omega \in \Omega} P^q(\omega)\)$. To connect notions, if $f$ is a nonnegative (non-zero) function on $\H$ with expectation $1$, then $P = \frac{f}{2^n}$ is a probability distribution, and $\log_2 \|f\|_q  = \frac{q-1}{q} \cdot \(n - H_q(X)\)$. Furthermore, $\log_2 \|f_{\e}\|_q = \frac{q-1}{q} \cdot \(n - H_q\(X \oplus Z\)\)$, where $X$ is a random variable on $\H$ distributed according to $P$ and $Z$ is an independent noise vector corresponding to a binary symmetric channel (BSC) with crossover probability $\e$. Hence, Proposition~\ref{pro:renyi entropy decrease} implies the following two claims
\begin{itemize}

\item Let $C$ be a pseudorandom code of rate $0 < R < 1$, and let $X$ be uniform on $C$ (note that $H_q(X) = Rn$). Then for any integer $q \ge 2$ and for any $0 \le \e \le \frac12$ such that $\l = \l(q,\e) \ge R$ holds
\[
H_q(X\oplus Z) ~\gtrsim~ \(1 - (\l - R)\) n.
\]

\item Let $Y$ be a random variable on $\H$ with $H_q(Y) = Rn$. Then for any $0 \le \e \le \frac12$ holds
\[
H_q(Y\oplus Z) ~\le~ \(1 - (\l - R)\) n.
\]

\end{itemize}

\noi This means that, up to lower order terms, the R\'enyi entropy of the random variable which is uniform on $C$ increases as fast as possible over the BSC.

\myblt Proposition~\ref{pro:renyi entropy decrease} shows that for any integer $q \ge 2$ and for all values of $\e$ for which $\l(q,\e)$ is greater than the rate of the code, the inequality in Theorem~\ref{thm:better-RE} is essentially tight for (characteristic functions of) pseudorandom linear codes, and hence for a typical linear code, since a random linear code is pseudorandom with high probability. Furthermore, as shown in \cite{S-old}, this inequality is tight for subcubes, which constitute a very special family of linear codes. This suggests, rather intriguingly, that it might be essentially tight for all linear codes. We point out that this is not the case. Let $n$ be even, and let $C = C^{\perp} = \big \{x = \(x_1,x_2\),x_1 = x_2\big\}$. Then we have the following claim, which we state somewhat informally at this point (see Section~\ref{sec:remaining} for the full statement).

\lem
\label{lem:BRE not tight}
Let $f = |C| \cdot 1_C$. Then for all $0 < \e < \frac12$ 
\[
\frac 1n \E_{T \sim \l} \log \| \E(f|T)\|_2 ~>~ \frac 1n \log_2 \|f_{\e}\|_2,
\]
where the strict inequality means that the LHS is larger than the RHS by an absolute constant, independent of $n$.
\elem

\noi With that, let us remark that the difference between the two sides of this inequality is upper bounded by a very small constant.

\noi {\it The probability of undetected error over a binary symmetric channel}

\noi Given a binary code $C \subseteq \H$, the undetected error probability of $C$ is the average probability (over the codewords) that a codeword transmitted over $BSC$ is distorted in such a way that the received word, though different from the transmitted one, also belongs to the code (\cite{Klove-Korzhik}). Let $C$ be a linear code, and let $\(a_0,...,a_n\)$ be its weight distribution. That is $a_k = \big |\left\{x \in C, |x| = k\right\} \big |$, for $k = 0,...,n$. Then the undetected error probability of $C$ can be expressed in terms of the weight distribution of $C$ and crossover probability $\e$:
\[
P_{\mathrm{ue}}(C,\e) ~=~ \frac{1}{|C|} \cdot \sum_{i=1}^n a_i(C) \e^i (1-\e)^{n-i}.
\]

\noi \noi The {\it best asymptotic undetected error exponent} for codes of rate $0 \le R \le 1$ and crossover probability $\e$ is defined as
\[
\pi_{\mathrm{ue}}(R,\e) ~=~ \mathrm{lim inf}_{n \rarrow \infty} \(-\frac 1n \min_C \log_2\(P_{\mathrm{ue}}(C,\e)\)\),
\]
where the minimum is taken over all codes $C \subseteq \H$ of cardinality $\lceil 2^{Rn} \rceil$. It is conjectured (see e.g., \cite{lev}) that the best error exponent is attained for random codes. This conjecture is known to hold (\cite{lits}) for a subset of the parameters $R$ and $\e$. This subset contains, in particular, the pairs $(R,\e)$ for which $1 + \log_2(1 - \e) \le R$, in which case it is known that $\pi_{\mathrm{ue}}(R,\e) = 1 - R$ (\cite{Klove-Korzhik}, Theorem 3.5.15).

\noi The following claim is a simple corollary of Proposition~\ref{pro:renyi entropy decrease}.

\cor
\label{cor:Becca-undetected error}
Let $0 < \e < \frac12$ and $0 < R < 1$ such that $1 + \log_2(1 - \e) \le R$. Let $C$ be a pseudorandom linear code of rate $R$. Then
\[
-\frac 1n \log_2\(P_{\mathrm{ue}}(C,\e)\) ~\gtrsim~ 1 - R.
\]

\ecor

\noi Some comments.

\myblt This result shows that pseudorandom codes of rate $R$ (and in particular Reed-Muller codes) have the best asymptotic undetected error exponent over $BSC$ with crossover probability $\e$, if $1 + \log_2(1 - \e) \le R$.

\myblt The claim of this corollary is an easy consequence of the case $q = \infty$ in Proposition~\ref{pro:renyi entropy decrease}. In fact, as observed e.g., in Lemma~1.4 in \cite{S}, if $f$ is the characteristic function of a linear code $C$, then $\|f_{\e}\|_{\infty} = f_{\e}(0) =  \sum_{i=0}^n a_i \e^i (1-\e)^{n-i} = (1-\e)^n + P_{\mathrm{ue}}(C,\e)$. The assumption that $1 + \log_2(1 - \e) \le R$ ensures that $(1-\e)^n$ does not dominate the undetected error probability if $C$ is pseudorandom, and hence that $\|f_{\e}\|_{\infty} \approx P_{\mathrm{ue}}(C,\e)$.

\noi {\it Rhombic $4$-tuples in random and pseudorandom codes}

\noi The statistics of possible inner distances in $k$-tuples of elements of a code, for some $k \ge 2$, is a central object of study in the best known approaches to bound the cardinality of error-correcting codes. In particular, the linear programming approach of Delsarte (\cite{dels, mrrw}) looks at the distance distribution of the code, that is the statistics of pairwise distances between codewords. The positive semidefinite approach of \cite{schrijver} (see also \cite{bgsv}) studies the geometry of a code in more detail, collecting the statistics of the possible ${k \choose 2}$-tuples of pairwise inner distances in all $k$-tuples of elements of a code, for some $k \ge 2$. In particular, statistics of inner distances of quadruples of codewords are studied in \cite{gms}. In a somewhat different direction, the extended linear programming approach suggested in \cite{cjj} studies inner distances in low-dimensional linear subspaces of a linear code.

\noi In this note we consider rhombic $4$-tuples of codewords, that is $4$-tuples $\(x,y,z,w\)$ with $|y-x| = |z-y| = |w-z| = |x-w|$. The relevance of counting such $4$-tuples in a code was discussed in \cite{llc}. For a linear code, counting $4$-tuples with $|y-x| = |z-y| = |w-z| = |x-w| = i$ is essentially equivalent to counting $4$-tuples $\(u_1,\ldots,u_4\)$ of codewords of weight $i$ which sum to $0$ (take $u_1 = x+y$, $u_2 = y+z$ and so on). We distinguish between 'trivial' $4$-tuples $\(u_1,\ldots,u_4\)$ with $u_1+...+u_4 = 0$ in which every vector appears an even number of times and 'non-trivial' $4$-tuples in which $u_1+...+u_4 = 0$ is the only linear dependence. We will focus on counting non-trivial $4$-tuples summing to $0$ in pseudorandom and random linear codes. Before describing our results, we introduce some notions and notation, following \cite{Lin-Mosh, LM}. In particular we use the random code model described in \cite{Lin-Mosh}.

\noi Let $h(t) = t \log_2\(\frac 1t\) + (1-t) \log_2\(\frac{1}{1-t}\)$ be the binary entropy function. Given $0 < \gamma \le \frac12$ and $0 < \l < h(\gamma)$, assume that an integer $n$ is such that $i = \gamma n$ and $\lambda n$ are integers. Let $L_i$ be the set of all vectors of weight $i$ in $\H$. Let $M$ be a random binary matrix with $\lambda n$ rows and $n$ columns. Let $C$ be a (random) linear code with parity check matrix $M$. As observed in \cite{Lin-Mosh}, with high probability the rate of $C$ is $R = 1 - \l$ and the number of vectors in $C \cap L_i$ is approximately $2^{(h(\gamma) - \l) \cdot n}$ (in particular, it is exponentially large). Let $X = \big |\{\(u_1,\ldots,u_4\) \in (C \cap L)^4:~u_1+...+u_4 = 0\}\big |$. As observed in \cite{Lin-Mosh, LM} (see e.g., the discussion following Theorem~1.2 in \cite{LM}),
\[
\frac 1n \log_2 \E X \approx \max\Big\{\psi(4,\gamma) + 2h(\gamma) - 3\l, 2h(\gamma) - 2\l \Big\},
\]
where the first term under the maximum on the RHS corresponds to the non-trivial $4$-tuples and the second term to the trivial ones \footnote{The number of trivial $4$-tuples summing to $0$ is, up to lower order terms, $|C \cap L_i|^2$.}. The function $\psi(4,\gamma)$ is a special case of $\psi(q,\gamma)$, which is a certain bivariate function defined and studied independently in \cite{Lin-Mosh} and \cite{KS2}. It is defined by the approximate identity $\big |\{\(u_1,\ldots,u_4\) \in L_i^4:~u_1+...+u_4 = 0\}\big | \approx 2^{(\psi(4,\gamma) + 2h(\gamma)) \cdot n}$. Observe that the expected number of non-trivial tuples is larger than that of the trivial ones iff $\psi(4,\gamma) > \l$.

\noi We proceed to describe our results. Let $N_i(C)$ denote the number of non-trivial $4$-tuples summing to $0$ in $C \cap L_i$.

\pro
\label{pro:tuples in random codes}

\noi Let $0 < R < 1$. Let $C \subseteq \H$ be a random linear code chosen according to the random code model described above with parameter $\l = 1 - R$. Let $0 < \gamma \le \frac12$ such that $h(\gamma) > \l$. Let $i = \gamma n$ (assuming $i$ is an integer). The following two cases hold  with probability tending to $1$ with $n$

\begin{enumerate}

\item $\psi(4,\gamma) < \l$. Then
\[
\frac 1n \log_2\(N_i\) ~<~ 2h(\gamma) - 2(1-R),
\]
where the strict inequality means that the RHS is larger than the LHS by an absolute constant.

\item $\psi(4,\gamma) > \l$.
\[
\frac 1n \log_2\(N_i\) ~\approx~ \psi(4,\gamma) + 2h(\gamma) - 3(1-R).
\]
\end{enumerate}

\epro

\noi Some comments.

\myblt As will be observed below, $\psi(4,\gamma) < h(\gamma)$ for $\gamma > 0$, and hence the condition $\psi(4,\gamma) > \l$ is stronger than $h(\gamma) > \l$. Note also that the condition $h(\gamma) > \l$ is both necessary and sufficient for $|C \cap L_i|$ to be exponentially large (with high probability).

\myblt The ratio of the total number of $4$-tuples in $C \cap L_i$ summing to $0$ to the number of trivial $4$-tuples with this property is $\frac{\|f\|_4^4}{\|f\|_2^4}$, where $f$ is the Fourier transform of the characteristic function of $C \cap L_i$. In particular, by \cite{KS2}, since the Fourier transform of $f$ is supported in $L_i$, this ratio is upperbounded (up to lower order terms) by $2^{\psi(4,\gamma) \cdot n}$. Observe that for  $h(\gamma) > \l$, the number of trivial $4$-tuples in $C \cap L_i$ is w.h.p. $\approx |C \cap L_i|^2 \approx 2^{2(h(\gamma) - \l) \cdot n}$. Hence for $\gamma$ such that $\psi(4,\gamma) > \l$ the second claim of the proposition implies that $\frac{\|f\|_4^4}{\|f\|_2^4} \lesssim 2^{(\psi(4,\gamma) - \l) \cdot n}$, obtaining a tighter estimate than that in \cite{KS2} in this case. If $\psi(4,\gamma) < \l$, the first claim of the proposition shows this ratio to be bounded by a constant, in accordance to the conjecture of \cite{Hastad-personal} presented in \cite{llc}. In the boundary case $\psi(4,\gamma) = \l$, it is easy to see from the proof of the proposition that the ratio $\frac{\|f\|_4^4}{\|f\|_2^4}$ is at most subexponential in $n$.

\pro
\label{pro:tuples in pseudorandom codes}

\noi Let $0 < R < 1$. Let $C \subseteq \H$ be a pseudorandom linear code of rate $R$. Let $0 < \gamma \le \frac12 $ and let $i = \gamma n$. There are two possible cases

\begin{enumerate}

\item $\gamma \ge 8^R + 1 - 3\sqrt{8^{2R-1} + 8^{R-1}}$. In this case, we have
\[
\frac 1n \log_2 N_i(C) ~\lesssim~ \psi(4,\gamma) + 2 h(\gamma) - 3(1-R).
\]

\item $\gamma < 8^R + 1 - 3\sqrt{8^{2R-1} + 8^{R-1}}$. In this case, we have
\[
\frac 1n \log_2 N_i(C) ~\lesssim~ -2\gamma \log_2\(\sqrt{8 + 8^{1-R}} - 3\).
\]

\end{enumerate}

\epro

\myblt It is easy to see that the function $f(R) = 8^R + 1 - 3\sqrt{8^{2R-1} + 8^{R-1}}$ decreases from $1/2$ to $0$ as $R$ goes from $0$ to $1$. It is also possible to verify that for $\gamma \ge f(R)$ we have $\psi(4,\gamma) > 1-R$. Hence Propositions~\ref{pro:tuples in random codes}~and~\ref{pro:tuples in pseudorandom codes} imply that if $C$ is a pseudorandom code and $i$ is sufficiently close to $n/2$, then the number of non-trivial $4$-tuples summing to $0$ in $C \cap L_i$ is bounded by the corresponding number in a random code of the same rate.
It seems instructive to compare this with the claim of Proposition~3.20 in \cite{HSS}, where a similar phenomenon is observed for the cardinality of the $i^{th}$ level of a pseudorandom code $C$, in comparison with that of a random code. Note that another way to describe $|C \cap L_i|$ is by counting the number of $2$-tuples summing to $0$ in $C \cap L_i$.

\noi This paper is organized as follows. We prove Proposition~\ref{pro:renyi entropy decrease} and Corollary~\ref{cor:Becca-undetected error} in Section~\ref{sec:Renyi entropy decrease}. Propositions~\ref{pro:tuples in random codes}~and~\ref{pro:tuples in pseudorandom codes} are proved in Section~\ref{sec:random-pseudorandom}. The remaining proofs appear in Section~\ref{sec:remaining}.

\section{Proofs of Proposition~\ref{pro:renyi entropy decrease} and Corollary~\ref{cor:Becca-undetected error}}
\label{sec:Renyi entropy decrease}

\subsection*{Proof of Proposition~\ref{pro:renyi entropy decrease}}

\noi We start with the first claim of the proposition. Let $C$ be a pseudorandom code of rate $0 < R < 1$, and let $f = \frac{2^n}{C} \cdot 1_C$. Let $q \ge 2$ be integer. Let $0 \le \e \le \frac12$ such that $\l = \l(q,\e) = 1 + \frac{1}{q-1} \cdot \log_2\(\e^q + (1-\e)^q\) \ge R$. Then
\[
\frac 1n \log \|f_{\e}\|_q ~\le~ \frac 1n \E_{T \sim \l} \log \| \E(f|T)\|_q ~=~  \frac{q-1}{q} \cdot \frac 1n \E_{T \sim \l} \(|T| - r_C(T)\) ~\le~
\]
\[
\frac{q-1}{q} \cdot \frac 1n  \(\E_{T \sim R} \(|T| - r_C(T)\) + (\l - R) n\) ~=~ \frac{q-1}{q} \cdot(\l - R) + o(1).
\]

\noi Here the first step is by Theorem~\ref{thm:better-RE}. The second step is by the fact observed in \cite{S} that if $C$ is a linear code then $\log_2 \| \E(f|T)\|_q = \frac{q-1}{q} \cdot \(|T| - r_C(T)\)$. The third step is a simple consequence of the Margulis-Russo formula (see e.g., the discussion in Remark~1.10 in \cite{S}) which implies that the function $m(\l) = \E_{T \sim \l} \big(|T| - r_{C}(T)\big)$ is increasing and that $m' \le n$. The fourth step follows from the definition of pseudorandom codes.

\noi We pass to the second claim of the proposition. Let $f$ be a nonnegative function on $\H$ with $\frac 1n \log_2 \|f\|_q = \frac{q-1}{q} \cdot (1-R)$. Let $\delta_x$ be the characteristic function of the point $x \in \H$. Observe that for any $x \in \H$, $\E \big(\(\delta_x\)_{\e}\big)^q = \E \big(\(\delta_0\)_{\e}\big)^q$. Since $\(\delta_0\)_{\e}(x) = \e^{|x|} (1-\e)^{n - |x|}$, $\E \big(\(\delta_0\)_{\e}\big)^q$ is easily computable, and it equals $2^{-n} \cdot \(\e^q + (1-\e)^q\)^n$. Taking this into account, we have
\[
\|f_{\e}\|_q^q ~=~ \E f_{\e}^q ~=~ \E \(\sum_{x \in \H} f(x) \(\delta_x\)_{\e}\)^q ~\ge~ \sum_{x \in \H} f^q(x) \E \big(\(\delta_x\)_{\e}\big)^q ~=~
\]
\[
2^{-n} \cdot \(\e^q + (1-\e)^q\)^n \cdot \sum_{x \in \H} f^q(x) ~=~ \(\e^q + (1-\e)^q\)^n \cdot \|f\|_q^q.
\]

\noi Hence
\[
\frac 1n \log_2 \|f_{\e}\|_q ~\ge~ \frac 1q \log_2\(\(\e^q + (1-\e)^q\)\) + \frac 1n \log_2 \|f\|_q ~=~
\]
\[
\frac 1q \log_2\(\(\e^q + (1-\e)^q\)\) + \frac{q-1}{q} \cdot (1-R) ~=~ \frac{q-1}{q} \cdot (\l-R).
\]

\noi It remains to deal with the case $q = \infty$. It is easy to see that this case requires no special treatment in the proof of the first claim of the proposition. The second claim can be proved for this case following the same outline as above. The only modification needed is replacing the inequality $\|f_{\e}\|_q^q \ge \sum_{x \in \H} f^q(x) \|\(\delta_x\)_{\e}\|_q^q$ with the inequality $\|f_{\e}\|_{\infty} \ge \|f\|_{\infty} \cdot \|\(\delta_x\)_{\e}\|_{\infty}$, where $x \in \H$ is the point for which $f(x) = \|f\|_{\infty}$.

\eprf

\subsection*{Proof of Corollary~\ref{cor:Becca-undetected error}}

\noi Let $0 < R < 1$. Let $C$ be a pseudorandom linear code of rate $R$. Let $f = \frac{2^n}{|C|} \cdot 1_C$. Let $\e_0$ be such that $1 + \log_2\(1 - \e_0\) = R$. Then for any $\e \ge \e_0$ (equivalently, $1 + \log_2(1 - \e) \le R$) holds
\[
\frac 1n \log_2 P_{\mathrm{ue}}(C,\e) ~\le~ (R-1) + \frac 1n \log_2 \|f_{\e}\|_{\infty} ~\le~  (R-1) + \frac 1n \log_2 \|f_{\e_0}\|_{\infty} ~\le~
\]
\[
(R-1) + \frac 1n \E_{T \sim R} \big(|T| - r_C(T)\big) ~\le~ R - 1 + o(1).
\]

\noi For the first step see the discussion following the statement of Corollary~\ref{cor:Becca-undetected error} in the introduction. The second step follows from the well-known fact that $\|f_{\e}\|_{\infty}$ decreases in $\e$ (this is a consequence of the semigroup property of noise operators). The third step follows from Theorem~\ref{thm:better-RE}, the definition of $\e_0$ and the fact that $\log_2 \|\E(f|T)\|_{\infty} = |T| - r_C(T)$. The last step follows from the definition of pseudorandom codes.

\eprf

\section{Proofs of Propositions~\ref{pro:tuples in random codes}~and~\ref{pro:tuples in pseudorandom codes}}
\label{sec:random-pseudorandom}

\subsection*{Proof of Proposition~\ref{pro:tuples in random codes}}

\prf

\noi  We will show that for any $0 < \gamma \le \frac12$ we have
\[
\Big | \frac 1n \log_2\(\E N_i\) -  \Big(\psi(4,\gamma) + 2h(\gamma) - 3\l\Big) \Big | ~\le~ o_n(1)
\]
and for $\gamma$ such that $\psi(4,\gamma) > \l$ we have $\sigma^2\(N_i\) \ll \E^2\(N_i\)$. More specifically we will show that in this case
\[
\frac 1n \log_2\(\sigma^2\(N_i\)\) ~<~ 2 \cdot \Big(\psi(4,\gamma) + 2h(\gamma) - 3\l\Big),
\]
where the strict inequality means that the RHS is larger than the LHS by an absolute constant.

\noi Note that this will imply both claims of the proposition. The first claim will follow by a first moment argument, since $N_i$ is a nonnegative random variable whose expectation for $\gamma$ such that $\psi(4,\gamma) < \l$ is exponentially smaller than $2^{2(h(\gamma) - \l) \cdot n}$. The second claim will follow by a second moment argument.

\noi In the following analysis we use the methods of \cite{Lin-Mosh, LM}, relying, in particular, on the properties of the function $\psi$ studied in \cite{Lin-Mosh,KS2}. We start with estimating the expectation of $N_i$. By the definition of the function $\psi$, the number of all $4$-tuples of vectors summing to $0$ in $L_i$ is given, up to lower order terms, by $2^{(\psi(4,\gamma) + 2h(\gamma)) \cdot n}$. Among these, the number of trivial $4$ tuples is about $2^{2h(\gamma) \cdot n}$, and hence (since $\psi(4,\gamma) > 0$ for $\gamma > 0$) it is negligible w.r.t. to the number of non-trivial ones. For a subset $S$ of $\H$, let $\mathrm{rank}(S)$ denote the dimension of the subspace spanned by the vectors in $S$. As observed in \cite{Lin-Mosh}, the probability that $S$ is contained in a random code $C$ is $2^{-\mathrm{rank}(S) \cdot \l n}$.
The rank of a non-trivial $4$-tuple is $3$, and therefore the probability that it is contained in a random code is $2^{-3\l n}$. The claim about $\E N_i$ follows.

\noi We pass to upper bounding the variance of $N_i$, when $\psi(4,\gamma) > \l$. Note that in this case $\psi(4,\gamma) + 2h(\gamma) > 3\l$, and hence $\E N_i$ is exponentially large. Let $U$ be the set of all non-trivial $4$-tuples in $L_i$. Then
\[
E N^2_i - {\E}^2 N_i ~=~ \sum_{S,T \in U} \Big(\mathrm{Pr}\{S \cup T \subseteq C\}  - \mathrm{Pr}\{S \in C\} \cdot \mathrm{Pr}\{T \in C\}\Big) ~\le~
\]
\[
\sum_{S,T \in U: \mathrm{rank}(S \cup T) < 6} \mathrm{Pr}\{S \cup T \subseteq C\} ~=~ \sum_{S,T \in U: \mathrm{rank}(S \cup T) < 6} 2^{-\mathrm{rank}(S \cup T) \cdot \l n}.
\]

\noi We proceed with some case analysis. The cardinality of $S \cup T$ is between $4$ and $8$. If $|S \cup T| = 4$ then $S = T$ and the rank of $S \cup T$ is $3$. Hence the contribution of this case to the summation on the RHS is $|U| \cdot 2^{-3\l n} = \E N_i \ll {\E}^2 N_i$. The case $|S \cup T| = 5$ is impossible, since if two non-trivial $4$-tuples summing to $0$ intersect in three places, they are identical. Consider the remaining cases. Note first that if $S \not = T$, then the vectors in $S \cup T$ satisfy at least two independent linear dependencies, and hence their rank is at most $|S \cup T| - 2$. Write $k$ for $|S \cup T|$ and $r$ for $\mathrm{rank}(S \cup T)$. The above discussion means that we need to consider cases in which $6 \le k \le 8$ and $3 \le r \le k-2$. Let $N(k,r)$ be the number of $k$-tuples $V \subseteq L_i$ of rank $r$. We will show in each of these cases that
\[
\frac 1n \log_2(N(k,r)) - r \l  ~<~ 2\psi(4,\gamma) + 4H(\gamma) - 6\l,
\]
where the strict inequality means that the RHS is larger than the LHS by an absolute constant. This will imply the required upper bound on $\sigma^2\(N_i\)$.

\noi The cases we need to consider are $k = 6$, $r \in \{3,4\}$; $k = 7$, $r \in \{3,4,5\}$; $k = 8$, $r \in \{4,5\}$. (Note that the case $k=8$ and $r=3$ is impossible, since all participating vectors are non-zero.)

\noi We proceed as in the proof of Proposition~3.3 in \cite{LM}, adhering to the notions and notation introduced in that paper. For $1 \le v \le \min\{r,k-r\}$, let $M(r,v) = \max_{\(a_1,...,a_v\)} \prod_{i=1}^v \E_{y_i} |K\(y_i\)|^{a_i}$, where the maximum is taken over all possible integer $v$-tuples $a_1,...,a_v$ which satisfy $a_d \ge 2$ for all $1 \le d \le v$ and $\sum_{d=1}^v a_d = r+v$. Here $K = K_i$ is the $i^{th}$ {\it Krawchouk polynomial}. The only fact we need to know about Krawchouk polynomials is that $\|K\|_a^a \approx \psi(a,\gamma) + \frac{a}{2} h(\gamma)$ (\cite{KS2}). Since $k$ in our case is constant and $1 \le v \le r \le k$, it follows from the argument in \cite{LM} that $N(k,r) \approx \max_{1 \le v \le \min\{r,k-r\}} M(r,v)$.

\noi Next, we claim that $\frac 1n \log_2(M(r,v)) \approx \psi(r-v+2,\gamma) + \frac{r+v}{2} \cdot h(\gamma)$. As in \cite{LM}, this follows easily from the following facts \cite{Lin-Mosh, KS2}: $\|K\|_a^a \approx \psi(a,\gamma) + \frac{a}{2} h(\gamma)$; $\psi(2,\gamma) = 0$, $\frac{\psi(a,\gamma)}{a-2}$ increases in $a$. Hence
\[
\frac 1n \log_2(N(k,r)) ~\approx~ \max_{1 \le v \le \min\{r,k-r\}} \Big\{\psi(r-v+2,\gamma) + \frac{r+v}{2} \cdot h(\gamma)\Big\} ~=~ \psi(r-v^{\ast}+2,\gamma) + \frac{r+v^{\ast}}{2} \cdot h(\gamma),
\]
where $v^{\ast} = \min\{r,k-r\}$. The second equality follows from the fact that $\frac{\partial \psi(a,\gamma)}{\partial a} < \frac{h(\gamma)}{2}$ (\cite{Lin-Mosh}).

\noi The only case in which $r > k - r$ is the case $k = 7$, $r = 3$. In this case $\frac 1n \log_2(N(k,r)) \approx \psi(2,\gamma) + 3h(\gamma) = 3 h(\gamma)$, and we need to verify (after simplification) that
\[
3\l ~<~ 2\psi(4,\gamma) + h(\gamma).
\]
This follows from the assumption $\l < \psi(4,\gamma)$. In fact, note that since $\psi(2,\gamma) = 0$ and $\frac{\partial \psi(a,\gamma)}{\partial a} < \frac{h(\gamma)}{2}$, we have $h(\gamma) > \psi(4,\gamma)$ and hence in our case $h(\gamma) > \l$ as well.

\noi In the remaining $6$ cases $r \ge k-r$, and hence $v^{\ast} = k-r$ and
\[
\frac 1n \log_2(N(k,r)) ~\approx~ \psi(2r-k+2,\gamma) + \frac k2 \cdot h(\gamma).
\]

\noi We now go over these cases and in each case state what is needed to be verified (after simplification):

\begin{enumerate}

\item $k = 6$, $r=3$: ~~~$3\l < 2\psi(4,\gamma) + h(\gamma)$.

\item $k = 6$, $r=4$: ~~~$2\l < \psi(4,\gamma) + h(\gamma)$.

\item $k = 7$, $r=4$: ~~~$\psi(3,\l) + 2\l < 2\psi(4,\gamma) + \frac12 h(\gamma)$.

\item $k = 7$, $r=5$: ~~~$\psi(5,\gamma) + \l < 2\psi(4,\gamma) + \frac12 h(\gamma)$.

\item $k = 8$, $r=4$: ~~~$2\l < 2\psi(4,\gamma)$.

\item $k = 8$, $r=5$: ~~~$\l < \psi(4,\gamma)$.

\end{enumerate}

\noi It is easy to see that all these inequalities follow from assumption $\l < \psi(4,\gamma)$ (and hence $\l < h(\gamma)$) and the following properties of $\psi$: $\psi(2,\gamma) = 0$ and $\frac{\partial \psi(a,\gamma)}{\partial a} < \frac{h(\gamma)}{2}$.

\eprf

\subsection*{Proof of Proposition~\ref{pro:tuples in pseudorandom codes}}

\noi We are going to prove a somewhat stronger claim, in that we are going to bound the total number of $4$-tuples summing to $0$ in $C \cap L_i$, rather than the number of non-trivial $4$-tuples with this property. Let $\tilde{N}_i$ be the number of all such $4$-tuples. From now on we bound $\tilde{N}_i(C)$ and hence, a fortiori, $N_i(C)$ for a pseudorandom code $C$.

\noi For a linear code $C$, let $f = |C| \cdot 1_{C^{\perp}}$. We claim that for any $0 \le \e \le \frac12$ holds $(1-2\e)^{4i}  \cdot \tilde{N}_i(C) \le \|f_{\e}\|_4^4$. To see this, we need some facts from Fourier analysis on the boolean cube \cite{O'Donnell}. Recall that $f = \sum_{x \in C} w_x$, where $\{w_x\}$ are the Walsh-Fourier characters, and furthermore that $f_{\e} = \sum_{x \in C} (1-2\e)^{|x|} w_x$. By the properties of Walsh-Fourier characters it follows that
\[
\|f_{\e}\|_4^4 ~=~ \sum_{x,y,z,w \in C^{\perp}: x+y+z+w=0}  (1-2\e)^{|x|+|y|+|z|+|w|} ~\ge~
\]
\[
\sum_{x,y,z,w \in C^{\perp}: x+y+z+w=0, |x| =|y| = |z| =|w| = i}  (1-2\e)^{4i} ~=~ (1-2\e)^{4i} \cdot \tilde{N}_i(C).
\]

\noi Now let $C$ be a pseudorandom linear code of rate $R$. In this case  $C^{\perp}$ is a pseudorandom code of rate $1-R$ (see the discussion following Definition~\ref{dfn:Becca}). Hence, $\E_{T \sim 1-R} \big(|T| - r_{C^{\perp}}(T)\big) \le o(n)$, and moreover, for any $\l \ge 1 - R$ we have $\E_{T \sim \l} \big(|T| - r_{C^{\perp}}(T)\big) \le \big(\l - (1-R)\big) \cdot n + o(n)$ (see the proof of Proposition~\ref{pro:renyi entropy decrease}).

\noi We apply Theorem~\ref{thm:better-RE} with $q=4$ to the function $f = |C| \cdot 1_{C^{\perp}}$. By the above discussion, for any $\e$ such that $\l(4,\e) \ge R$ holds
\[
\tilde{N}_i(C) ~\le~ \frac{\|f_{\e}\|_4^4}{(1-2\e)^{4i}} ~\le~ 2^{o(n)} \cdot \frac{2^{\(\l_4(\e) - (1-R)\) \cdot n}}{(1-2\e)^{4i}}.
\]

\noi Passing to exponents and recalling that $i = \gamma n$ gives
\[
\frac1n \log_2\(\tilde{N}_i(C)\) ~\lesssim~ \min_{\e: \l(4,\e) \ge 1 - R} \Big\{3 (\l_4 - (1-R)) - 4 \gamma \log_2\(1-2\e\) \Big\}.
\]

\noi Next, we investigate the function $g(\e) = 3 \l(4,\e) - 4 \gamma \log_2\(1-2\e\)$. We need the following auxiliary lemma.

\lem
\label{lem:lambda-psi-4}

\noi For any any $0 < \gamma < \frac12$ holds
\[
\min_{0 \le \e \le \frac12} ~g(\e)  ~=~  \psi(4,\gamma) + 2 h(\gamma).
\]
Furthermore, the function $g$ is unimodal on $[0,1/2]$, decreasing to its point of global minimum $\e_0$ and increasing afterwards, where $\e = \e_0$ is determined by $\gamma = \(\frac12 - \e\) \cdot \frac{(1-\e)^3 - \e^3}{(1-\e)^4 + \e^4}$.
\elem

\noi This lemma is proved in Section~\ref{sec:remaining}. We assume its validity and proceed with the proof of the proposition. Note that $\l(4,\e)$ decreases from $1$ to $0$ as $\e$ goes from $0$ to $1/2$. Let $\e_1$ be determined by $\l\(4,\e_1\) = 1 - R$. There are two possible cases: if $\e_1 \ge \e_0$ then
\[
\frac1n \log_2\(\tilde{N}_i(C)\) ~\lesssim~ \min_{\e: \l(4,\e) \ge 1 - R} \big\{g(\e)\big\} -3(1-R) ~=~ \min_{0 \le \e \le \frac12} \big\{g(\e)\big\} -3(1-R)  ~=~
\]
\[
\psi(4,\gamma) + 2 h(\gamma) - 3(1-R).
\]
Otherwise, we get
\[
\frac1n \log_2\(\tilde{N}_i(C)\) ~\lesssim~  \min_{\e: \l(4,\e) \ge 1 - R} \big\{g(\e)\big\} -3(1-R) ~=~ g\(\e_1\) - 3(1-R) ~=~ -4\gamma \log_2\(1 - 2\e_1\).
\]

\noi To complete the proof of the proposition, we verify that $\(1-2\e_1\)^2 = \sqrt{8 + 8^{1-R}} - 3$ and that $\e_1 \ge \e_0$ iff $\gamma \ge 8^R + 1 - 3\sqrt{8^{2R-1} + 8^{R-1}}$. Let $y_0 = \(1 - 2\e_0\)^2$ and $y_1 = \(1 - 2\e_1\)^2$. For any $\e$ and $y = (1-2\e)^2$ holds $8\(\e^4 + (1-\e)^4\) = y^2 + 6y + 1$. Hence $y_1$ is determined by $y_1^2 + 6y_1 + 1 = 8^{1-R}$, or equivalently $y_1 = \sqrt{8 + 8^{1-R}} - 3$.

\noi The condition $\e_1 \ge \e_0$ is equivalent to $y_1 \le y_0$. For any $\e$ and $y = (1-2\e)^2$ holds
$\(\frac12 - \e\) \cdot \frac{(1-\e)^3 - \e^3}{(1-\e)^4 + \e^4} = \frac{y^2 + 3y}{y^2+6y+1}$. Hence $y_0$ is determined by $\gamma = \frac{y^2 + 3y}{y^2+6y+1}$, or equivalently
\[
y_0 ~=~ \frac{6 \gamma - 3 + \sqrt{(3-6\gamma)^2 + 4\gamma(1-\gamma)}}{2(1-\gamma)}.
\]

\noi Given this, it is not hard to verify that the condition $y_1 \le y_0$ is in fact equivalent to $\gamma \ge 8^R + 1 - 3\sqrt{8^{2R-1} + 8^{R-1}}$ (we omit the details).

\eprf

\section{Remaining proofs}
\label{sec:remaining}

\subsection*{Proof of Lemma~\ref{lem:BRE not tight}}
\noi We start with a more detailed statement of the lemma.

\lem
\label{lem:full-BRE not tight}
Let $n$ be even, and let $C = C^{\perp} = \big \{x = \(x_1,x_2\),x_1 = x_2\big\}$. Let $f = |C| \cdot 1_C$. Then for $0 \le \e \le \frac12$ we have
\[
\frac 1n \log_2 \|f_{\e}\|_2^2 ~=~ \frac{1}{2} \cdot \log_2\(1 + (1-2\e)^4\),
\]
and
\[
\frac 1n \E_{T \sim \l} \log \| \E(f|T)\|^2_2 ~=~ \frac12 \cdot \(1 + \log_2\(\e^2 + (1-\e)^2\)\)^2.
\]

\noi Furthermore, $\log_2\(1 + (1-2\e)^4\) < \(1 + \log_2\(\e^2 + (1-\e)^2\)\)^2$ for all $0 < \e < \frac12$.
\elem

\prf

\noi As noted in the proof of Proposition~\ref{pro:tuples in pseudorandom codes}, we have (recalling $C = C^{\perp}$) that $f_{\e} = \sum_{x \in C} (1-2\e)^{|x|} w_x$. Hence, by Parseval's identity,
\[
\|f_{\e}\|_2^2 ~=~ \sum_{x \in C} (1-2\e)^{2|x|} ~=~ \sum_{\mathrm{even}~i,~i=0}^n {{n/2} \choose {i/2}} (1-2\e)^{2i} ~=~
\]
\[
\sum_{j=0}^{n/2} {{n/2} \choose j} (1 - 2\e)^{4j} ~=~ \(1 + (1-2\e)^4\)^{n/2},
\]
proving the first claim of the lemma.

\noi Next,  $T \subseteq [n]$ decomposes naturally into $\(T_1,T_2\)$ and $r_C(T) = |T_1 \cup T_2|$. Hence, for $\l = \l(2,\e) = 1 + \log_2\(\e^2 + (1-\e)^2\)$, we have
\[
\E_{T \sim \l} \log \| \E(f|T)\|^2_2 ~=~ \l n - \E_{T \sim \l} r_C(T) ~=~ \l n - \E_{T \sim \l} |T_1 \cup T_2| ~=~ \l n - \E_{T \sim \l} \Big(|T_1| + |T_2| - |T_1 \cap T_2|\Big) ~=~
\]
\[
\E_{T \sim \l} |T_1 \cap T_2| ~=~ \E_{R \sim \l^2} |R| ~=~ \frac{n}{2} \cdot \l^2,
\]
proving the second claim of the lemma. In the last step, note that $R$ is a random subset of $[n/2]$.

\noi Let $y = (1-2\e)^2$. Then the third claim of the lemma amounts to proving $\log_2^2(1+y) > \log_2\(1 + y^2\)$ for $0 < y < 1$. Let $f(y) = \log_2(1+y)$. Then we need to show that $f^2(y) > f\(y^2\)$. Define $g(z) = \ln\(f\(e^z\)\) = \ln\(\log_2\(1 + e^z\)\)$, for $-\infty < z \le 0$. Then we need to show that $2g(z) > g(2z)$ for $z < 0$. Since $g(0) = 0$ this would follow if we show that $g'' < 0$. In fact, it is easy to see that, up to a positive factor, $g''(z)$ is proportional to $\ln\(1+e^z\) - 1 < 0$, completing the proof of the third claim and the proof of the lemma.

\eprf

\subsection*{Proof of Lemma~\ref{lem:lambda-psi-4}}
\noi We prove a more general result.

\lem
\label{lem:lambda-psi-general}

\noi For any $q \ge 2$ and any $0 < \gamma < \frac12$ holds
\[
\min_{0 \le \e \le \frac12} \Big\{(q-1) \l(q,\e) - q\gamma \log_2(1-2\e)\Big\}  ~=~  \psi(q,x) + \frac q2 h(\gamma).
\]
Furthermore, the function $g(\e) = (q-1) \l(q,\e) - q\gamma \log_2(1-2\e)$ is unimodal on $[0,1/2]$, decreasing to its point of global minimum $\e_0$ and increasing afterwards, where $\e = \e_0$ is determined by $\gamma = \(\frac12 - \e\) \cdot \frac{(1-\e)^{q-1} - \e^{q-1}}{\(1-\e\)^q + \e^q}$.
\elem

\prf
Recalling that $\l(q,\e) = 1 + \frac{1}{q-1} \log_2\(\e^q + (1-\e)^q\)$, we have $g(\e) = (q-1) + \log_2\(\e^q + (1-\e)^q\) - q \gamma \log_2(1-2\e)$. After some simplification, we get
\[
g' ~\sim~ \gamma - \(\frac12 - \e\) \cdot \frac{(1-\e)^{q-1} - \e^{q-1}}{\(1-\e\)^q + \e^q}.
\]

\noi As shown in Lemma~2.5 in \cite{KS2}, the function $a(p,\e) = \(\frac12 - \e\) \cdot \frac{(1-\e)^{p-1} - \e^{p-1}}{\(1-\e\)^p + \e^p}$ is decreasing from $\frac12$ to $0$ as $\e$ goes from $0$ to $\frac12$. It follows that $g$ is unimodal, and that its unique minimum is attained at $\e_0$. By Proposition~2.6 in \cite{KS2}, the value of this minimum is $g\(\e_0\) ~=~ \psi(q,x) + \frac q2 h(\gamma)$.

\eprf

\end{document}